\documentstyle[12pt,epsfig]{article}

\voffset     = - 0.10 in
\begin{document}
\vspace*{1cm}

\begin{center}{\Large\bf
Non-compensation of the Barrel Combined Calorimeter Prototype}
\end{center}
\vspace{2cm}
\begin{center}

{\Large\bf  Y.A.~Kulchitsky, M.V.~Kuzmin}

\bigskip

{\it 
Institute of Physics, National Academy of Sciences, Minsk, Belarus} \\

\smallskip

\& \ \  
{\it JINR, Dubna, Russia}
\end{center}

\vspace*{\fill}

\begin{abstract}
 The e/$\pi$ ratio for the Barrel Combined Calorimeter Prototype,
composed from electromagnetic LArg calorimeter and hadronic Tile
calorimter was investigated. Response of Combined Calorimeter 
on pions and electrons in the energy region 20 -- 300 GeV was studied. 
Found $e/h = 1.37 \pm 0.01 \pm 0.02$ is in good agreement with results
from previous Combined Calorimeter test but has more precisions.
\end{abstract}
\newpage

\section{Introduction}

ATLAS project \cite {ccATLAS} represents a general-purpose detector to 
investigate pp collisions in the energy region up to 14 TeV. 
Designed barrel part of calorimetry system consists of the 
electromagnetic liquid argon (Larg) 
calorimeter, using the accordion geometry and   
a large scintillating tile  hadronic  barrel
calorimeter, based on  a  sampling technique using 
steel absorber material 
and scintillating plates read  out by wavelength shifting fibers.  
Detailed description of them can be found in \cite{LARG},\cite{TDR}.
 
In that work we report the results 
on the studying of the e/$\pi$ and e/h for the Combined calorimeter, 
composed from Larg and TILE prototypes. 
A e/h is a characteristic number of any calorimeter system and  describe 
the non-compensation of calorimeter response on hadrons relatively 
to electrons. As electro-magnetic calorimeter is slim for 
hadrons, great part of hadron shower is outside from LArg. Therefore e/pi for 
Combined calorimeter is of special interest. 

That investigation was performed 
on the basis of data on exposure of ATLAS Barrel Combined Calorimeter
Prototype  
in beams of pions and electrons 
with energy 20 -- 300 GeV in April 1996.  
Results on the studying e/h for TILE
calorimeter can be found in \cite{budagov96-72}.

\section{The Electromagnetic Calorimeter}

  The electro-magnetic LArg calorimeter  prototype 
consists of a stack of three azimuthal modules, each one spanning 
$9^\circ$ in azimuth and extending  over 2~m along the {\em z} 
direction.  The calorimeter structure is defined by  
2.2~mm thick steel-plated lead absorbers, folded to an accordion shape and 
separated by 3.8~mm gaps, filled with liquid argon; the signals are collected 
by Kapton electrodes  located in the gaps.

The calorimeter extends from an inner radius of 131.5~cm to 
an outer radius of 182.6~cm, representing (at $\eta$ = 0) a total
of 25 radiation lengths ($X_0$), or 1.22 interaction lengths 
($\lambda$) for protons. The calorimeter is longitudinally
segmented  into three compartments of 
$9~X_0$, $9~X_0$ and $7~X_0$, respectively.
The $\eta \times \phi$ segmentation is $0.018 \times 0.02$ 
for the first two longitudinal compartments and  $0.036 \times 0.02$ 
for the last compartment, where $\eta=-\log(\tan \frac{\theta}{2}) $. Each
read-out cell has full projective geometry 
in $\eta$ and in $\phi$. \
 The calorimeter was located inside 
a large cylindrical cryostat with 2~m internal diameter, filled with liquid 
argon. 

\section{The Hadronic Calorimeter}

The hadron calorimeter prototype consists of an azimuthal
stack of five modules. 
Each module covers $2\pi/64$ in azimuth and extends 
1~m along the {\em z} direction, such that the front face  covers 
$100\times20$~cm$^2$. The radial depth, 
from an inner radius of 200~cm to an outer radius of 380~cm, 
 accounts for   8.9~$\lambda$ at $\eta$~=~0 (80.5~$X_0$) for protons.  
Read-out cells are defined by grouping together a bundle of fibers 
into one photo-multiplier (PMT). 
Each of the 100 cells is read out by two PMTs and is fully
projective in azimuth (with $\Delta \phi =2\pi/64 \approx 0.1$),
while the segmentation along the {\em z}
axis is made by grouping fibers into read-out cells spanning
$\Delta z = 20$~cm ($\Delta \eta \approx 0.1$) and is therefore not projective.
Each module is read out in four longitudinal segments 
(corresponding to about 1.5, 2, 2.5 and 3~$\lambda$ at $\eta$~=~0). 
More details of this prototype can be found in
\cite{ccATLAS},\cite{budagov96-72}. 

The beam incident angle was, as in the previous combined run, of about 
11$^0$, but now the impact point was 8 cm left from the center to avoid side
leakage.

\section{Experimental Setup}

\begin{figure}[ht]
\vspace{-0.5cm}
 \begin{center}
  \mbox{\epsfig{figure=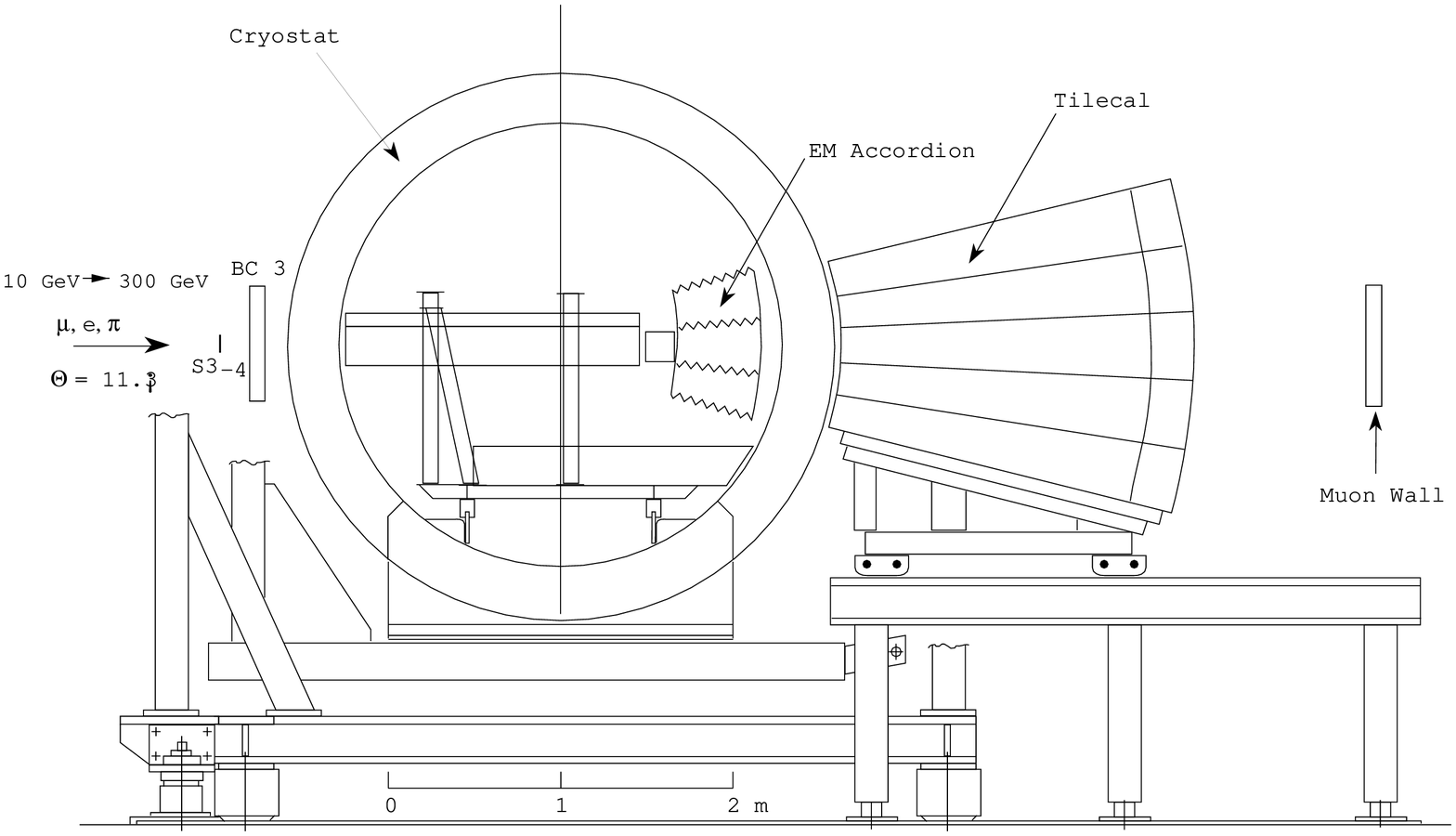,width=0.95\textwidth,height=0.4\textheight}}
 \end{center}
 \caption{ 
   Ttest beam setup  for the combined LArg and tile
   calorimeter combined run
\label{fig:setup}}
\end{figure}

To simulate the ATLAS setup, the Tile calorimeter was 
placed on a fixed table, just behind the LArg Accordion cryostat, as shown in
figure~\ref{fig:setup}.

To optimize the containment of hadronic showers
the electro-magnetic calorimeter was located as close as possible 
to the back of the cryostat as in the previous combined run.
Early showers in the liquid argon were kept to a minimum by placing
light foam material in the cryostat upstream of the calorimeter. 

With respect to previous combined test beam setup
\cite{94test}, a new element is present.In order to try to 
understand the energy loss in dead material
between the active part of the LArg detector and the Tilecal, 
a layer of scintillator called the mid-sampler was installed . 
The mid-sampler consisted of five scintillators, 20~cm $\times$ 
100~cm each, fastened directly to the
front face of the Tilecal modules.  The scintillator was 1~cm thick, and was
readout using ten 1 mm WLS fibers on each of the long sides.

Beam quality and geometry were monitored with a set of beam chambers 
and trigger hodoscopes placed upstream of the LArg cryostat. 
The momentum bite of the beam was always less than 0.5\%.  
Single-track pion events were selected 
offline by requiring the pulse height of the beam scintillation counters (S3-4 on the picture)
and the energy released in the presampler of the electro-magnetic calorimeter
to be compatible with that of a single particle.   
Beam halo events were removed with appropriate cuts on the horizontal 
and  vertical positions of the incoming track impact point
as measured with the two beam chambers (BC on the picture).

For this layout the effective distance between the two active parts of 
the detector is of the order of 50~cm, instead of the 25~cm as 
foreseen in the ATLAS setup. The amount of material has been 
quantified to be about $2 X_0$
in between the two calorimeters. This value is similar to the ATLAS
design value, but the material type is different: steel instead of
aluminum for the cryostat. The total depth corresponds to about 
10.1 $\lambda$, to be compared with the 9.6 foreseen in the ATLAS 
setup \cite{ccATLAS}. A large scintillator wall (``muon wall'')
covering about 1~$m^2$ of surface has been placed on
the back of the calorimeter to quantify  leakage.

\section{Reconstruction of Electron Energy}

To separate electrons from the muons, hadrons and events 
with interactions in dead material before combined calorimeter 
the following cuts were applied:

\begin{itemize} 
\item 
events with only physical trigger were selected;
\item 
cuts on beam geometry (signals from wire chambers BC1, BC2, BC3 were used);
\item 
cuts on signal from scintillator counters S1, S2, S3, S4 
and pressampler to reject events with interactions before combined 
calorimeter;
\item 
to separate electrons from hadrons cuts on responses of Tile calorimeter 
samplings were applied;
\item 
cuts on total energy deposition to reject muons.
\end{itemize}

\begin{figure*}[tbph]
\begin{center}   
\begin{tabular}{c}
\epsfig{figure=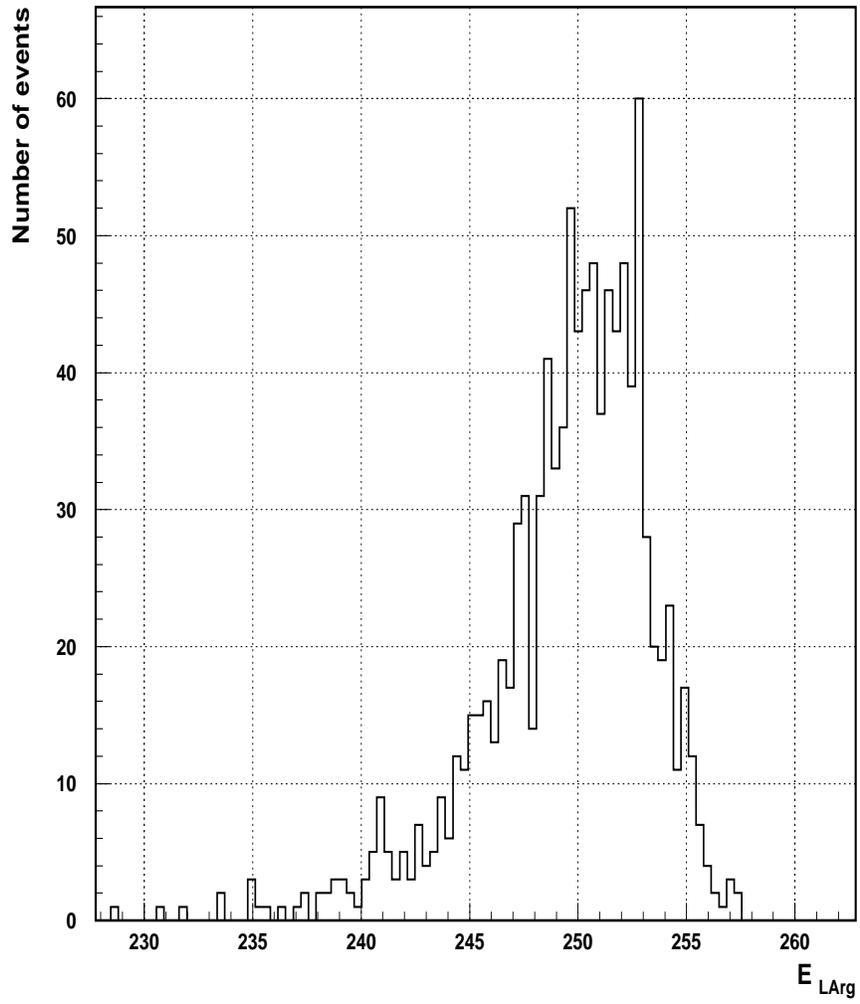,width=0.95\textwidth,height=0.8\textheight} 
\\
\end{tabular}
\end{center}
       \caption{
         Response of the electro-magnetic calorimeter (LArg) on 287,5 GeV 
         electrons.
\label{f001}}
\end{figure*}

The characteristic feature of the electro-magnetic shower is its small 
transverse radius with the comparison of hadronic one. 
For reconstruction of electron energy cluster in $3\times7$
 cells of electro-magnetic calorimeter was used to avoid including noise 
in total response. Meanwhile, direct response of LArg on electrons is 
not Gaussian-like, as shown on Fig.\ \ref{f001}.

\begin{figure*}[tbph]
\begin{center}   
\begin{tabular}{c}
\epsfig{figure=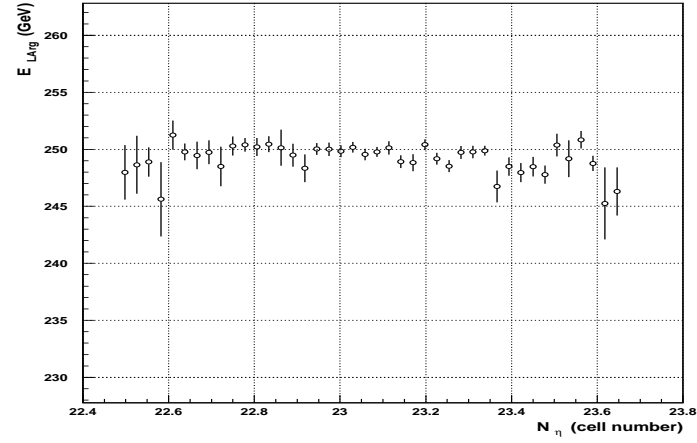,width=0.75\textwidth,height=0.35\textheight}  
\\
\epsfig{figure=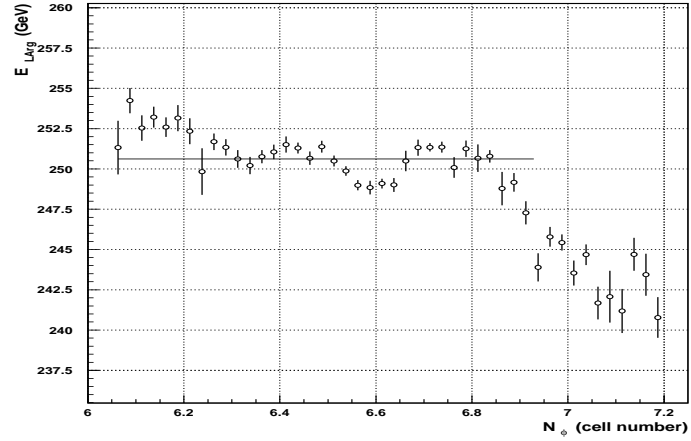,width=0.75\textwidth,height=0.35\textheight}
 \\
\end{tabular}
\end{center}
       \caption{ Top:
         Dependence of average LArg response on 287.5 GeV electrons 
         on $\eta$ position of impacting electron 
         as a function of $\eta$-coordinate. 
         $\eta$-coordinate expressed
         in LArg cell numbers along $\eta$ direction.
                Bottom:
        Dependence of average LArg response on 287.5 GeV electrons  
         on $\phi$ position  of impacting electron, as a function of
         $\phi$-coordinate. 
         Mean value of the response was extracted
         from fitting the spectrum with a line.
         $\phi$-coordinate expressed
         in LArg cell numbers along $\phi$ direction.
       \label{f002}}
\end{figure*}

 Explanation of such behavior appears to be dependence of response on an 
impact point of electron, which is shown on  Fig.\ \ref{f002}, where
 $\eta$ dependence (top picture) can be considered negligible in terms 
of uncertainties, there 
$\eta$-coordinate is expressed in cell numbers of electro-magnetic calorimeter
along $\eta$ direction.
For along $\phi$ (bottom picture) one can observe strong influence of LArg 
internal structure on total electron response, where $\phi$ also in cells
numbers. Therefore total response of LArg on electrons represents a 
sum of normal distributions with different mean values, according to 
their statistical weights, and shouldn't be Gaussian-like. 

As there is no any strong dependence of calorimeter response on the electron 
impacting point along $\eta$, 
therefore  to achieve the mean value of energy spectrum
$\phi$-dependence was fitted with the line
in the same $\phi$ region for all energies. 
To put mean value of LArg response on electrons  
correspondent to known beam energy scale factor 
$\alpha = 1.166 \pm 0.003$ was applied. 
Received mean values of responses, and reconstructed energy of electrons 
for 20 --  287.5 GeV beams are gathered in Table\ \ref{T1}.

Direct error of the fitting parameter was considered as statistical error. 
An systematic error 0.4\% was introduced to achieve total $\chi^2 = 1$ for 
all set of energy points.
Additional error in 0.3\% was introduced due to the 
uncertainty of scale factor. 
Achieved linearity of LArg response on electrons is in terms of $\pm 1.5
\%$, which is comparable with results from Ref.\ \cite{larg97-72}.

\subsection{The e/$\pi$ Ratio}

To extract $e/\pi$ ratio for combined calorimeter system, 
one should go to the absolute energy scale for each calorimeter. 
In our case for electrons the signal from LArg calorimeter is sufficient 
for energy reconstruction, 
so to receive response of Combined calorimeter on pions, 
and extract $e/\pi$ ratio the following formula was used:
\begin{equation}
\frac{e}{\pi} = 
\frac{e_{L\ (3\times7)}\ E_{em}^e}{e_{L\ (11\times11)}\ E_{em}^\pi + e_T E_{h}^\pi + e_{cr} E_{cryo}^{\pi}}\ ,
\label{e07}
\end{equation}
where 
$E_{em}^e$ and $E_{em}^\pi$ are response of LArg calorimeter on electrons
and pions,
$e_{L\ (11\times11)}= 1.1$ is calibration constant for 
LArg calorimeter for window $(11\times11)$, 
 $e_{L\ (3\times7)}=1.166$ is scale factor for 
 electron response in Larg window $(3\times7)$,
$E_{h}^\pi$ is response of Tile calorimeter on pions,
$e_T = k /(e/ \pi)_T = 0.145$ 
is calibration constant for Tile calorimeter and 
$k = 300 (GeV)/E_{h}^\pi = 0.156$,
$E_{cryo}^\pi$ is the energy loss in the cryostat,
$e_{cr} = e_T\ c/a$ is the calibration constant for cryostat.
For the case of stand-alone calorimeter that formula leads us to the 
obvious expression $e/\pi = E^e/E^\pi$.
Calculated $e/\pi$ ratios for beam energies 20 -- 300 are gathered in 
Table \ref{T2}.

For the case of 300 GeV point, offset in 12.5 GeV was added to the 
reconstructed mean value of response on 287.5 GeV electrons.
To achieve good $\chi^2$ of fitting an additional error for 20 GeV 
point 2\% was introduced due to the large uncertainty 
in definition of response on electrons (beam spot in the $6.5< \phi <7.2$).
The $e/h$ ratio was extracted from received data by fitting them 
with expression \cite{Wigmans}: 
\begin{equation}
\frac{e}{\pi} = \frac{e/h}{1 + (e/h - 1)\ 0.11\ lnE}\ .
\label{e08}
\end{equation}

\begin{figure*}[tbph]
     \begin{center}
\epsfig{figure=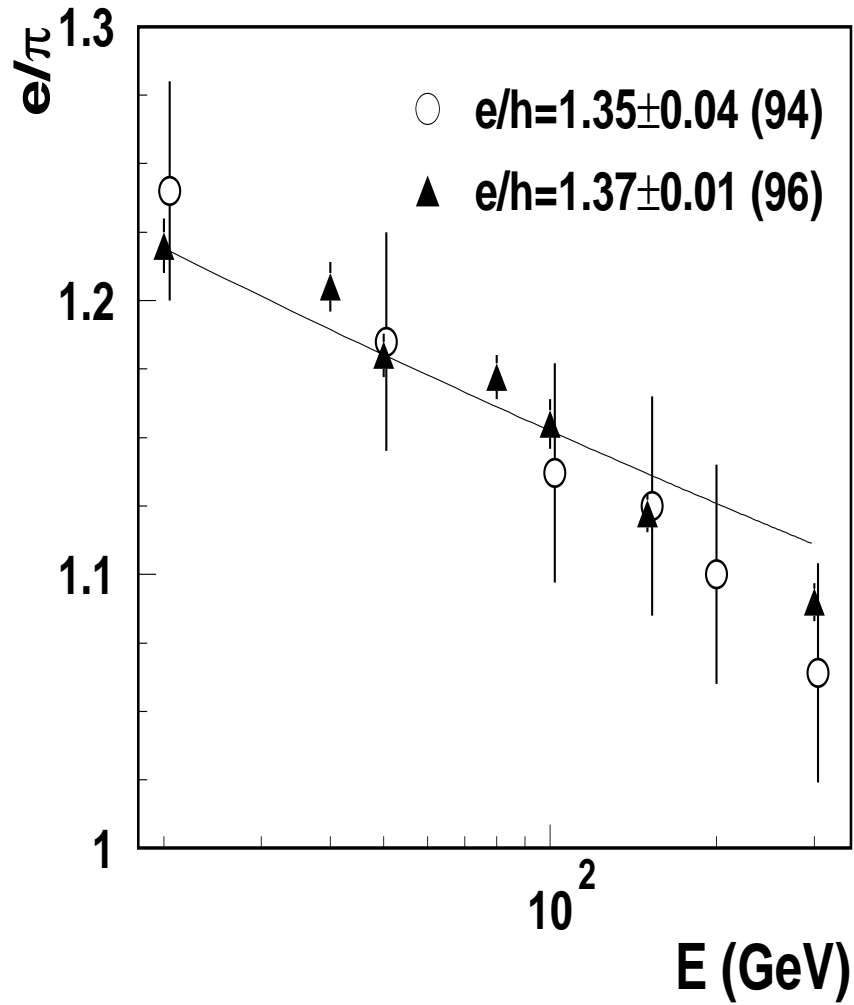,width=0.95\textwidth,height=0.85\textheight}
     \end{center}
     \caption{
       Distribution of the $e/ \pi$ ratio versus the beam energy, 
       fitted by the expression (\ref{e08}). 
       Black circles represent 
       our results. 
       Open circles represent
       results from Ref.\ \cite{94test}.
       \label{f006}}
\end{figure*}

Fig.\ \ref{f006} show the $e/\pi$ ratios for Combine 96 (black circles)
and Combine 94 (open circles). 
The solid curve is fit of our data by function (\ref{e08}).
As the result of fit we found $e/h = 1.37 \pm 0.01 \pm 0.02$, which is in a 
good agreement with data, received in 1994 ($e/h = 1.35 \pm 0.04$), 
but more precision. 
With the comparison for stand-alone prototype of Tile 
calorimeter \cite{budagov96-72} $e/h = 1.23 \pm 0.02$ for incident particle
angle $10^\circ$, we received larger value for $e/h$. 
That could be explained if LArg calorimeter is more uncompensated then 
Tile calorimeter,
as $e/h$ for Combined setup represents a average on $e/h$ for 
both calorimeters.

\section{Conclusion}

To measure e/$\pi$ and $e/h$ ratio for prototype of ATLAS Barrel 
Combined Calorimeter responses on pions and electrons were studied. 
Found $e/h = 1.37 \pm 0.01 \pm 0.02$ is in good agreement with results
\cite{94test} but more precisions.

\section{Acknowledgments}

This work is the result of the efforts of many people 
from ATLAS Collaboration. the authors are greatly 
indebted to all Collaboration for their test beam setup and data taking.
Authors are grateful for M.\ Nessi  
and J.\ Budagov for their attention and support of this work. 
We are indebted to M.\ Cobal, D.\ Constanzo, 
B.\ Lung-Jensen and V.V.\ Vinogradov 
for the valuable discussions and constgructive advices.




\begin{table}[tbph]
\begin{center}
\caption{
  Mean value of Larg response on electrons for the various beam energy.
\label{T1}}
\begin{tabular}{|c|c|c|}  
\hline
Beam energy & Mean value of response   & Reconstructed      \\
 (GeV) &       (GeV)                   &  energy  (GeV)     \\ \hline
 287.5 & 250.6  $\pm$ 0.30 $\pm$ 1.00  & 292.3 $\pm$ 2.0    \\ \hline 
 150   & 130.9  $\pm$ 0.12 $\pm$ 0.52  & 152.7 $\pm$ 0.9    \\ \hline 
 100   & 86.36  $\pm$ 0.08 $\pm$ 0.34  & 100.7 $\pm$ 0.6    \\ \hline 
 80    & 69.09  $\pm$ 0.05 $\pm$ 0.28  & 80.6  $\pm$ 0.4   \\ \hline 
 50    & 42.45  $\pm$ 0.04 $\pm$ 0.17  & 49.5  $\pm$ 0.3   \\ \hline 
 40    & 33.84  $\pm$ 0.03 $\pm$ 0.14  & 39.5  $\pm$ 0.2   \\ \hline 
 20    & 16.88  $\pm$ 0.03 $\pm$ 0.07  & 19.7  $\pm$ 0.1   \\ \hline 
\end{tabular}
\end{center}
\end{table}

\begin{table}[tbph]
\begin{center}
\caption{ 
  e/$\pi$ ratio of Combined Calorimeter for the various beam energy.
\label{T2}}
\begin{tabular}{|c|c|}
\multicolumn{2}{l}{\mbox{~~~}}\\
\hline
$E_{beam}$ (GeV)  & e/$\pi$ ratio \\ 
\hline
300 & 1.114 $\pm$ 0.013 \\ 
\hline 
150 & 1.137 $\pm$ 0.013 \\ 
\hline 
100 & 1.158 $\pm$ 0.012 \\ 
\hline   
80  & 1.170 $\pm$ 0.013 \\ 
\hline 
50  & 1.174 $\pm$ 0.012 \\ 
\hline 
40  & 1.186 $\pm$ 0.012 \\ 
\hline 
20  & 1.278 $\pm$ 0.016 \\ 
\hline 
\end{tabular}
\end{center}
\end{table}



\newpage

\clearpage

\end{document}